\documentclass[12pt]{article}
\usepackage{color}
\usepackage{amssymb,amsmath}
\usepackage[dvipdfmx]{graphicx}

\newcommand{\bea}{\begin{eqnarray}}
\newcommand{\eea}{\end{eqnarray}}
\begin{document}
\setlength{\baselineskip}{0.7cm}
\setlength{\baselineskip}{0.7cm}
\begin{titlepage} 
\begin{flushright}
OCU-PHYS 457 \\
YGHP17-03 
\end{flushright}
\vspace*{10mm}
\begin{center}{\Large\bf Fermion Dark Matter 
in Gauge-Higgs Unification}
\end{center}
\vspace*{10mm}
\begin{center}
{\large Nobuhito Maru}$^{a}$, 
{\large Takashi Miyaji}$^{a}$, 
{\large Nobuchika Okada}$^{b}$, 
and 
{\large Satomi Okada}$^{c}$
\end{center}
\vspace*{0.2cm}
\begin{center}
${}^{a}${\it 
Department of Mathematics and Physics, Osaka City University, \\ 
Osaka 558-8585, Japan}
\\[0.2cm]
${}^{b}${\it Department of Physics and Astronomy, University of Alabama, \\
Tuscaloosa, Alabama 35487, USA} 
\\[0.2cm]
${}^{c}${\it Graduate School of Science and Engineering, Yamagata University, \\
Yamagata 990-8560, Japan} 
\end{center}
\vspace*{2cm}
\begin{abstract} 

We propose a Majorana fermion dark matter in the context of a simple gauge-Higgs Unification (GHU) scenario 
   based on the gauge group $SU(3) \times U(1)^\prime$ in 5-dimensional Minkowski space 
   with a compactification of the 5th dimension on $S^1/Z_2$ orbifold. 
The dark matter particle is identified with the lightest mode in $SU(3)$ triplet fermions 
   additionally introduced in the 5-dimensional bulk. 
We find an allowed parameter region for the dark matter mass around a half of 
   the Standard Model Higgs boson mass, which is consistent with the observed dark matter density 
   and the constraint from the LUX 2016 result for the direct dark matter search.   
The entire allowed region will be covered by, for example, the LUX-ZEPLIN dark matter experiment in the near future. 
We also show that in the presence of the bulk $SU(3)$ triplet fermions 
   the 125 GeV Higgs boson mass is reproduced through the renormalization group evolution 
   of Higgs quartic coupling with the compactification scale of around $10^8$ GeV. 

\end{abstract}
\end{titlepage}

\section{Introduction}

The existence of dark matter (DM) is promising from the various cosmological observations 
  and one of the keys for exploring physics beyond the Standard Model (SM). 
It is still a mystery in particle physics and cosmology to clarify the identities of the dark matter particle. 
Among various possibilities, the so-called Weakly Interacting Massive Particle is a prime candidate 
  for the DM particle, which is the thermal relic from the early Universe 
  and whose relic abundance is calculable independently of the history of the Universe 
  before the DM has gotten in thermal equilibrium. 
A variety of experiments aiming for directly/indirectly detecting DM particles is ongoing and planned, 
  and the discovery of the dark matter may be around the corner. 
In this paper we consider a fermion DM in the context of a simple gauge-Higgs Unification (GHU) scenario 
  in 5-dimensions and identify a model-parameter region which is consistent with the current experimental 
  constraints.

The GHU scenario \cite{GH} is a unique candidate for new physics beyond the SM, 
  which offers a solution to the gauge hierarchy problem without invoking supersymmetry. 
An essential property of the GHU scenario is that the SM Higgs doublet is identified 
   with an extra spatial component of the gauge field in higher dimensions. 
Associated with the higher-dimensional gauge symmetry, 
   the GHU scenario predicts various finite physical observables, 
   irrespective of the non-renormalizability of the scenario, 
   such as the effective Higgs potential~\cite{1loopmass, 2loop},  
   the effective Higgs coupling with digluon/diphoton \cite{MO, Maru, diphoton}, 
   the anomalous magnetic moment $g-2$~\cite{g-2}, 
   and the electric dipole moment~\cite{EDM}.

In the previous paper by some of the present authors \cite{diphoton}, 
   the one-loop contributions of Kaluza-Klein (KK) modes 
   to the Higgs-to-digluon and Higgs-to-diphoton couplings were calculated 
   in a 5-dimensional GHU model by introducing color-singlet bulk fermions 
   with a half-periodic boundary condition, in addition to the SM fermions.  
It was shown that the color-singlet bulk fermions play a crucial role 
   not only to explain the observed Higgs-to-digluon and Higgs-to-diphoton couplings,  
   but also to achieve the 125 GeV Higgs boson mass. 
See also Ref.~\cite{MOcolor} for extended analysis including color-triplet bulk fermions. 
As a bonus, 
  it was pointed out that the lightest KK mode of the bulk fermions can be a DM candidate 
  by choosing their hypercharges appropriately. 
The main purpose of this paper is to pursue this possibility and investigate the DM physics 
  in the context of the GHU scenario. 
For related works on the DM physics in GHU scenarios, see Ref.~\cite{GHDM}.

Towards the completion of the GHU scenario as new physics beyond the SM,  
   we need to supplement a DM candidate to the scenario. 
In order to keep the original motivation of the GHU scenario to solve the gauge hierarchy 
   problem, the DM candidate to be introduced must be a fermion.  
Since the GHU scenario is defined in a higher dimensional space-time 
  with a gauge group into which the SM gauge group is embedded, 
  it would be the most natural/general to introduce a DM candidate as a bulk fermion 
  of a certain representation under the gauge group of the GHU scenario.  
Hence, the DM candidate is accompanied by its partners in decomposition 
  of the SM gauge groups and has a Yukawa coupling with the SM Higgs doublet, 
  which originates from the higher-dimensional gauge interaction.  
Thanks to the structure of the  GHU scenario, once the representation of the bulk fermion 
  is defined, the Yukawa coupling is predicted. 
This is in a sharp contrast with 4-dimensional DM models, where Yukawa couplings 
  are generally undetermined.   
In addition, as we will show in Sec.~4, the fermion DM multiplet in the bulk 
  plays a crucial role to lower the compactification scale of the (5-dimensional) GHU scenario 
  while reproducing the observed Higgs boson mass of 125 GeV.

The plan of this paper is as follows. 
In the next section, we consider a 5-dimensional GHU model 
   based on the gauge group $SU(3)\times U(1)^\prime$ 
   with an orbifold $S^1/Z_2$ compactification. 
In this context, we propose a Majorana fermion DM scenario,  
   where a DM particle is provided as the lightest mass eigenstate 
   in a pair of bulk $SU(3)$ triplet fermions introduced in the bulk 
   along with a bulk mass term and a periodic boundary condition.    
In Sec.~3, we focus on the case that the DM particle communicates with the SM particles  
   through the Higgs boson. 
Solving the Boltzmann equation, we identify an allowed parameter region 
   of the model to reproduce the observed DM density. 
In Sec.~4, we further constrain the allowed parameter region by considering 
   the upper limit of the elastic scattering cross section of the DM particle off with nuclei 
   from the current DM direct detection experiments. 
An effective field theoretical approach of the GHU scenario 
   will be discussed in Sec.~5, and the 125 GeV Higgs boson mass is reproduced 
   in the presence of the bulk $SU(3)$ triplet fermions with certain boundary conditions. 
The compactification scale is determined in order to reproduce the Higgs boson mass of 125 GeV. 
The last section is devoted to conclusions.

\section{Fermion DM in GHU}
We consider a GHU model based on the gauge group 
  $SU(3) \times U(1)^\prime$ \cite{SSS} in a 5-dimensional flat space-time 
  with orbifolding on $S^1/Z_2$ with radius $R$ of $S^1$. 
In our setup of bulk fermions including the SM fermions, we follow Ref.~\cite{CCP}: 
  the up-type quarks except for the top quark, 
  the down-type quarks and the leptons are embedded, respectively, into ${\bf 3}$, $\overline{{\bf 6}}$, 
  and ${\bf 10}$ representations of $SU(3)$. 
In order to realize the large top Yukawa coupling, the top quark is embedded 
  into a rank $4$ representation of $SU(3)$, namely $\overline{{\bf 15}}$. 
The extra $U(1)^\prime$ symmetry works to yield the correct weak mixing angle, 
  and the SM $U(1)_Y$ gauge boson is realized  
 by a linear combination between the gauge bosons of the $U(1)^\prime$ 
  and the $U(1)$ subgroup in $SU(3)$ \cite{SSS}. 
Appropriate $U(1)^\prime$ charges for bulk fermions are assigned 
  to yield the correct hyper-charges for the SM fermions.

The boundary conditions should be suitably assigned 
   to reproduce the SM fields as the zero modes. 
While a periodic boundary condition corresponding to $S^1$ 
  is taken for all of the bulk SM fields, 
  the $Z_2$ parity is assigned for gauge fields and fermions 
  in the representation ${\cal R}$ 
 by using the parity matrix $P={\rm diag}(-,-,+)$ as
\bea
A_\mu (-y) = P^\dag A_\mu(y) P, \quad A_y(-y) =- P^\dag A_y(y) P,  \quad 
\psi(-y) = {\cal R}(P) \gamma^5 \psi(y) 
\label{parity}
\eea 
  where the subscripts $\mu$ ($y$) denotes the four (the fifth) dimensional component. 
With this choice of parities, the $SU(3)$ gauge symmetry is explicitly broken down to $SU(2) \times U(1)$. 
A hypercharge is a linear combination of $U(1)$ and $U(1)^\prime$ in this setup. 
One may think that the $U(1)_X$ gauge boson which is orthogonal to the hypercharge $U(1)_Y$ 
  also has a zero mode. 
However, the $U(1)_X$ symmetry is anomalous in general and broken at the cutoff scale and hence,  
  the $U(1)_X$ gauge boson has a mass of order of the cutoff scale \cite{SSS}. 
As a result, zero-mode vector bosons in the model are only the SM gauge fields.

Off-diagonal blocks in $A_y$ have zero modes because of the overall sign in Eq.~(\ref{parity}), 
  which corresponds to an $SU(2)$ doublet. 
In fact,  the SM Higgs doublet ($H$) is identified with 
\bea
A_y^{(0)} = \frac{1}{\sqrt{2}}
\left(
\begin{array}{cc}
0 & H \\
H^\dag & 0 \\
\end{array}
\right). 
\eea
The KK modes of $A_y$ are eaten by KK modes of the SM gauge bosons 
 and enjoy their longitudinal degrees of freedom 
 like the usual Higgs mechanism.

The parity assignment also provides the SM fermions as massless modes, 
  but it also leaves exotic fermions massless. 
Such exotic fermions are made massive by introducing 
  brane localized fermions with conjugate $SU(2) \times U(1)$ charges 
  and an opposite chirality to the exotic fermions, 
  allowing us to write mass terms on the orbifold fixed points. 
In the GHU scenario, the Yukawa interaction is unified 
  with the gauge interaction, so that the SM fermions 
  obtain the mass of the order of the $W$-boson mass 
  after the electroweak symmetry breaking. 
To realize light SM fermion masses, one may introduce $Z_2$-parity odd bulk mass terms 
  for the SM fermions, except for the top quark. 
Then, zero mode fermion wave functions with opposite chirality are localized 
  towards the opposite orbifold fixed points and as a result, 
  their Yukawa couplings are exponentially suppressed by the overlap integral of the wave functions. 
In this way, all exotic fermion zero modes can be heavy and the small Yukawa couplings for the light SM fermions 
  can be realized by adjusting the bulk mass parameters. 
In order to realize the top quark Yukawa coupling, 
  we introduce a rank $4$ tensor representation, namely, 
  a symmetric $\overline{{\bf 15}}$ without a bulk mass \cite{CCP}. 
This leads to a group theoretical factor $2$ enhancement 
  of the top quark mass as $m_t = 2 m_W$ at the compactification scale \cite{SSS}. 
Note that this mass relation is desirable since the top quark pole mass receives QCD threshold 
  corrections which push up the mass about 10 GeV.

Now we discuss the DM sector in our model. 
In addition to the bulk fermions corresponding to the SM quarks and leptons, 
   we introduce a pair of extra bulk fermions $\psi, \tilde{\psi}$ 
   which are triplet representations under the bulk $SU(3)$ and have a $U(1)^\prime$ charge $1/3$.  
With this choice of the $U(1)^\prime$ charge, 
   the triplet bulk fermions include electric-charge neutral components 
   and a linear combination among the charge neutral components serves as the DM particle.    
Associated with $S^1$ we impose the periodic boundary condition in the fifth dimension, 
   while the $Z_2$ parity assignments are chosen as 
\bea
\psi(-y) = P \gamma^5 \psi(y), \quad 
\tilde{\psi} = - P \gamma^5 \tilde{\psi}(y). 
\eea 
After the electroweak symmetry breaking, the lightest mass eigenstate among the bulk triplets 
  is identified with the DM particle. 
As we will discuss in Sec.~\ref{Hmass}, these bulk fermions also play a crucial role 
  to reproduce the observed Higgs boson mass of 125 GeV.

The Lagrangian relevant to our DM physics discussion is given by
\bea
{\cal L}_{{\rm DM}} = \overline{\psi} \; i D\!\!\!\!/ \; \psi + \overline{\tilde{\psi}} \;  i D\!\!\!\!/  \; {\tilde \psi} 
  - M(\overline{\psi} \tilde{\psi} + \overline{\tilde{\psi}} \psi) 
  + \delta(y) \left[ \frac{m}{2} \overline{\psi_{3R}^{(0)c}} \psi_{3R}^{(0)} 
  +\frac{\tilde{m}}{2} \overline{\tilde{\psi}_{3L}^{(0)c}} \tilde{\psi}_{3L}^{(0)} 
  + {\rm h.c.} \right], 
\label{DMLagrangian}
\eea
 where the covariant derivative and a pair of the bulk $SU(3)$ triplets are given by
\bea
&&D\!\!\!\!/ = \Gamma^M (\partial_M - ig A_M -ig' A'_M), \\
&&\psi = (\psi_1, \psi_2, \psi_3)^T, \quad 
\tilde{\psi} = (\tilde{\psi}_1, \tilde{\psi}_2, \tilde{\psi}_3 )^T. 
\eea
With the non-trivial orbifold boundary conditions, the bulk $SU(3)$ triplet fermions are 
   decomposed into the SM $SU(2)$ doublet and singlet fermions. 
As we will see later, the DM particle is provided as a linear combination of 
   the second and third components  of the triplet fermions. 
In Eq.~(\ref{DMLagrangian}) we have introduced a bulk mass ($M$) to avoid exotic massless fermions.  
Here we have also introduced Majorana mass terms on the brane at $y=0$ 
   for the zero-modes of the third components of the triplets ($\psi_{3R}^{(0)}$ and $\tilde{\psi}_{3L}^{(0)}$),   
   which are singlet under the SM gauge group. 
The superscript ``c" denotes the charge conjugation. 
With the Majorana masses on the brane, the DM particle in 4-dimensional effective theory 
   is a Majorana fermion, and hence its spin-independent cross section with nuclei  
   through the $Z$-boson exchange vanishes in the non-relativistic limit.

Let us focus on the following terms in Eq.~(\ref{DMLagrangian}), 
   which are relevant to the mass terms in 4-dimensional effective theory: 
\bea
{\cal L}_{{\rm mass}} &=& 
 \overline{\psi} i \Gamma^5 (\partial_y -ig \langle A_y \rangle) \psi 
 + \overline{\tilde{\psi}} i \Gamma^5 (\partial_y -ig \langle A_y \rangle) \tilde{\psi} 
 - M(\overline{\psi} \tilde{\psi} + \overline{\tilde{\psi}} \psi) 
 \nonumber \\
 &&+ \delta(y) \left[ \frac{m}{2} \overline{\psi_{3R}^{(0)c}} \psi_{3R}^{(0)} 
 +\frac{\tilde{m}}{2} \overline{\tilde{\psi}_{3L}^{(0)c}} \psi_{3L}^{(0)} 
 + {\rm h.c.} \right], 
\label{L_mass}
\eea
 where $\Gamma^5= i \gamma^5$. 
Expanding the bulk fermions in terms of KK modes as
\bea
&&\psi(x,y) = \frac{1}{\sqrt{2\pi R}} \psi^{(0)}(x) 
+ \frac{1}{\sqrt{\pi R}} \sum_{n=1}^\infty \psi^{(n)}(x) \cos \left(\frac{n}{R} y\right)~({\rm for}~\psi_{1L,2L,3R}, \tilde{\psi}_{1R,2R,3L}), \\
&&\psi(x,y) = \frac{1}{\sqrt{\pi R}} \sum_{n=1}^\infty \psi^{(n)}(x) \sin \left(\frac{n}{R} y\right)~({\rm for}~\psi_{1R,2R,3L}, \tilde{\psi}_{1L,2L,3R}), 
\eea
and integrating out the fifth coordinate $y$, we obtain the expression in 4-dimensional effective theory.  
The zero-mode parts for the electric-charge neutral fermions are found to be 
\bea
{\cal L}^{{\rm zero-mode}}_{{\rm mass}} 
&=& 
 i m_W \left( 
 \overline{\psi_{2L}^{(0)}} \psi_{3R}^{(0)} + \overline{\tilde{\psi}_{3L}^{(0)}} \tilde{\psi}_{2R}^{(0)} \right) 
 -M \left(\overline{\psi_{2L}^{(0)}} \tilde{\psi}_{2R}^{(0)} + \overline{\tilde{\psi}_{3L}^{(0)}} \psi_{3R}^{(0)} 
\right) + {\rm h.c.} 
\nonumber \\
&& +
\frac{m}{2} \overline{\psi_{3R}^{(0)c}} \psi_{3R}^{(0)} 
 + \frac{\tilde{m}}{2} \overline{\tilde{\psi}_{3L}^{(0)c}} \tilde{\psi}_{3L}^{(0)} 
 + {\rm h.c.} 
\nonumber \\
& \to & 
 - m_W \left( 
 \overline{\psi_{2L}^{(0)}} \psi_{3R}^{(0)} - \overline{\tilde{\psi}_{3L}^{(0)}} \tilde{\psi}_{2R}^{(0)} \right) 
 -M \left(\overline{\psi_{2L}^{(0)}} \tilde{\psi}_{2R}^{(0)} + \overline{\tilde{\psi}_{3L}^{(0)}} \psi_{3R}^{(0)} 
\right) + {\rm h.c.} 
\nonumber \\
&& -
\frac{m}{2} \overline{\psi_{3R}^{(0)c}} \psi_{3R}^{(0)} 
 - \frac{\tilde{m}}{2} \overline{\tilde{\psi}_{3L}^{(0)c}} \tilde{\psi}_{3L}^{(0)} 
 + {\rm h.c.} 
\eea
where $m_W=gv/2$ is the $W$-boson mass, and the arrow means the phase rotations 
$\psi_{3R}^{(0)} \to i \psi_{3R}^{(0)}$ and $\tilde{\psi}_{3L}^{(0)} \to i \tilde{\psi}_{3L}^{(0)}$. 
It is useful to rewrite these mass terms in a Majorana basis defined as  
\bea
&&\chi \equiv \psi^{(0)}_{3R} + \psi^{(0)c}_{3R}, \quad 
\tilde{\chi} \equiv \tilde{\psi}^{(0)}_{3L} + \tilde{\psi}^{(0)c}_{3L}, 
\nonumber \\
&&\omega \equiv \psi^{(0)}_{2L} + \psi^{(0)c}_{2L}, \quad 
\tilde{\omega} \equiv \tilde{\psi}^{(0)}_{2R} + \tilde{\psi}^{(0)c}_{2R}, 
\eea
and we then express the mass matrix (${\cal M}_{{\rm N}}$) as 
\bea
{\cal L}_{{\rm mass}}^{{\rm zero-mode}} 
 &=& 
-\frac{1}{2}(
\begin{array}{cccc}
\overline{\chi} & \overline{\tilde{\chi}} & \overline{\omega} & \overline{\tilde{\omega}} \\
\end{array}
)
{\cal M}_{{\rm N}}
\left(
\begin{array}{c}
\chi \\
\tilde{\chi} \\
\omega  \\
\tilde{\omega }\\
\end{array}
\right)  \nonumber \\
&=&
 -\frac{1}{2}(
\begin{array}{cccc}
\overline{\chi} & \overline{\tilde{\chi}} & \overline{\omega} & \overline{\tilde{\omega}} \\
\end{array}
)
\left(
\begin{array}{cccc}
m      & M            &  m_W  & 0 \\
M      & {\tilde m} &   0       & -m_W \\
m_W &     0        &   0       &  M \\
0       & -m_W    &    M      & 0 \\
\end{array}
\right)
\left(
\begin{array}{c}
\chi \\
\tilde{\chi} \\
\omega  \\
\tilde{\omega }\\
\end{array}
\right) .
\label{MassMatrix}
\eea
The zero-modes of the charged fermions, $\psi_{1L}^{(0)}$ and  $\tilde{\psi}_{1R}^{(0)}$, 
   have a Dirac mass of $M$.

To simplify our analysis, we set $m=\tilde{m}$, and in this case we find a simple expression 
   for the mass eigenvalues of $M_N$ as 
\bea
&&m_1=\frac{1}{2} \left( m - \sqrt{4 m_W^2 + (m-2M)^2} \right), \nonumber \\
&&m_2=\frac{1}{2} \left( m + \sqrt{4 m_W^2 + (m-2M)^2} \right), \nonumber \\
&&m_3=\frac{1}{2} \left( m - \sqrt{4 m_W^2 + (m+2M)^2} \right) , \nonumber \\
&&m_4=\frac{1}{2} \left( m + \sqrt{4 m_W^2 + (m+2M)^2} \right),  
\label{MassEigenvalues}
\eea 
for the mass eigenstates defined as 
 $  (\chi \; \tilde{\chi} \; \omega \;  \tilde{\omega})^T = U_{\cal M} \; (\eta_1 \; \eta_2 \;  \eta_3 \;  \eta_4)^T $  
 with a unitary matrix 
\bea
&&
U_{\cal M}=
\left(
\begin{array}{cccc}
 u_1 &   u_2 & u_3       & u_4  \\
-u_1 &  -u_2 & u_3        & u_4  \\
 1             & 1                                         & 1 &   1 \\
 1             & 1                                         & -1 & -1 \\
\end{array}
\right)
\left(
\begin{array}{cccc}
 \frac{1}{c_1} &  0  &  0  & 0 \\
   0  &  \frac{1}{c_2}  &  0  & 0 \\
  0  &  0  &  \frac{1}{c_3}  & 0 \\
  0  &  0  &  0  &  \frac{1}{c_4} \\
\end{array}
\right), 
\eea 
where 
\bea 
\label{u_i}  
  u_1 &=& \frac{m_1-M}{m_W},  \quad  u_2 = \frac{m_2-M}{m_W},  \quad  
  u_3 = \frac{m_3+M}{m_W},  \quad  u_4 = \frac{m_4 + M}{m_W},  \\ 
  c_1 &=& \sqrt{2 (u_1^2 +1)}, \quad  c_2= \sqrt{2 (u_2^2 +1)}, \quad  
  c_3 = \sqrt{2 (u_3^2 +1)}, \quad c_4= \sqrt{2 (u_4^2 +1)}. 
\label{c_i}
\eea
Note that without loss of generality we can take $M, m \geq 0$.   
Considering the current experimental constraints from the search for an exotic charged fermion, 
   we may take $M \gtrsim 1~{\rm TeV} \gg m_W$ \cite{PDG}. 
In this case, the lowest mass eigenvalue (dark matter mass $m_{\rm DM}$) is given by $|m_1|$. 
 From the explicit form of the mass matrix ${\cal M}_{{\rm N}}$ in Eq.~(\ref{MassMatrix}) and $M \gg m_W$, 
   we notice two typical cases for the constituent of the DM particle: 
   (i) the DM particle is mostly an SM singlet when $m=\tilde{m} \lesssim M$, or 
   (ii) the DM particle is mostly a component in the SM $SU(2)$ doublets when $m=\tilde{m} \gtrsim M$.    
In the case (i), the DM particle communicates with the SM particle essentially through the SM Higgs boson.   
On the other hand, the DM particle is quite similar to the so-called Higgsino-like neutralino DM 
   in the minimal supersymmetric SM (MSSM) for the case (ii).
Since the Higgsino-like neutralino DM has been very well-studied in many literatures,\footnote{
In this case, a pair of the DM particles mainly annihilates into the weak gauge bosons 
  through the SM $SU(2)$ gauge coupling, and the observed DM relic abundance 
  can be reproduced with the DM mass of around 1 TeV~\cite{ ArkaniHamed:2006mb}.   
}
   we focus on the case (i) in this paper. 
Note that the case (i) is a realization of the so-called Higgs-portal DM from the GHU scenario. 
We emphasize that in our scenario, the Yukawa couplings in the original Lagrangian 
   are not free parameters, but are the SM $SU(2)$ gauge coupling,  
   thanks to the structure of the GHU scenario.

Now we describe the coupling between the DM particle and the Higgs boson. 
In the original basis, the interaction can be read off from Eq.~(\ref{MassMatrix}) 
   by $v \to v+h$ as 
\bea
{\cal L}_{{\rm Higgs-coupling}} &=&  
-\frac{1}{2} 
\left( \frac{m_W}{v} \right) h 
\left(
\begin{array}{cccc}
\overline{\chi} & \overline{\tilde{\chi}} & \overline{\omega} & \overline{\tilde{\omega}} \\
\end{array}
\right)
\left(
\begin{array}{cccc}
0       &   0     &  1     & 0 \\
0       &   0     &   0       & - 1 \\
1       &   0    &   0       &  0 \\
0       &  -1    &    0     & 0 \\
\end{array}
\right)
\left(
\begin{array}{c}
\chi \\
\tilde{\chi} \\
\omega  \\
\tilde{\omega }\\
\end{array}
\right)  \nonumber \\
&& =
-\frac{1}{2} 
\left( \frac{m_W}{v} \right) h 
\left(
\begin{array}{cccc}
\overline{\eta_1} & \overline{\eta_2} & \overline{\eta_3} & \overline{\eta_4} \\
\end{array}
\right)
{\cal C}_{h}
\left(
\begin{array}{c}
\eta_1 \\
\eta_2 \\
\eta_3 \\
\eta_4\\
\end{array}
\right),  
\label{Hint}
\eea
where $h$ is the physical Higgs boson, and the explicit form of the matrix ${\cal C}_h$ is given by 
\bea
{\cal C}_h  \equiv 
\left(
\begin{array}{cccc}
{\cal C}_{1} & {\cal C}_{5} & 0 & 0 \\
{\cal C}_{5} & {\cal C}_{2} & 0 & 0 \\
0 & 0 & {\cal C}_{3}& {\cal C}_{6} \\
0 & 0 & {\cal C}_{6}& {\cal C}_{4} \\
\end{array}
\right), 
\label{Hphys} 
\eea
where 
\bea
{\cal C}_{1} = \frac{4 u_1}{c_1^2}, \; \;
{\cal C}_{2} = \frac{4 u_2}{c_2^2}, \; \;
{\cal C}_{3} = \frac{4 u_3}{c_3^2}, \;  \;
{\cal C}_{4} = \frac{4 u_4}{c_4^2}, \;  \;
{\cal C}_{5} = \frac{2 (u_1+ u_2)}{c_1 c_2}, \; \;
{\cal C}_{6} = \frac{2 (u_3+ u_4)}{c_3 c_4}. 
\eea 
The interaction Lagrangian relevant to the DM physics is given by 
\bea
{\cal L}_{{\rm DM-H}} = 
  -\frac{1}{2} \left( \frac{m_W}{v} \right) {\cal C}_1 \; h \; \overline{\psi_{{\rm DM}}} \; \psi_{{\rm DM}}  
  -\frac{1}{2} \left(\frac{m_W}{v} \right)  {\cal C}_5 \; h \left( \overline{\eta_2} \; \psi_{{\rm DM}} + {\rm h.c.} \right),    
\label{DMHcoupling}
\eea
where we have identified the lightest mass eigenstate $\eta_1$ as the DM particle ($\psi_{\rm DM}$).

\section{Dark Matter Relic Abundance}
\label{RA}
In this section, we evaluate the DM relic abundance and identify an allowed parameter region 
  to be consistent with the Planck 2015 measurement of the DM relic density \cite{Planck2015} (68 \% confidence level):  
\bea
\Omega_{{\rm DM}} h^2 = 0.1198 \pm 0.0015. 
\eea
In our model, the DM physics is controlled by only two free parameters, namely, $m$ and $M$. 
As we discussed in the previous section, we focus on the Higgs-portal DM case 
  with $0 \leq m \lesssim M$. 
Using $M \gg  m_W$, we can easily derive approximate formulas for parameters involved 
  in our DM analysis. 
For the mass eigenvalues listed in Eq.~(\ref{MassEigenvalues}), we find 
\bea
 m_1 \simeq -M + m - \frac{m_W^2}{2 M-m},  \quad   
 m_2 \simeq   M +  \frac{m_W^2}{2 M-m}.  
\eea
By using these formulas, we express $u_{1, 2}$ and $c_{1, 2}$ in Eqs.~(\ref{u_i}) and (\ref{c_i}) as 
\bea 
 u_1 \simeq - \frac{2 M-m}{m_W}, \; \; u_2 \simeq \frac{m_W}{2 M -m},  \; \;
 c_1 \simeq  \sqrt{2} \left(\frac{2 M-m}{m_W}\right),  \;\; c_2 \simeq \sqrt{2}, 
\eea
which lead to 
\bea 
  {\cal C}_1 \simeq - \frac{2 m_W}{2M -m}, \; \; \; {\cal C}_5 \simeq 1. 
\eea
For $m \lesssim M$ and a fixed value of $M \gg m_W$, 
   the DM particle can be light when $m \simeq M$, otherwise $m_{\rm DM} \simeq M$
   while $m_2 \simeq M$ for any values of $m \lesssim M$. 
The coupling of a DM particle pair with the Higgs boson is always suppressed by $|{\cal C}_1| \ll 1$ 
   while ${\cal C}_5 \simeq 1$.

According to the interaction Lagrangian in Eq.~(\ref{DMHcoupling}), we consider two main annihilation processes 
  of a pair of DM particles. 
One is through the $s$-channel Higgs boson exchange, and 
  the other is the process $\psi_{\rm DM} \psi_{\rm DM} \to h h$ through 
  the exchange of $\eta_2$ in the $t/u$-channel. 
Since $|{\cal C}_1| \ll 1$ and ${{\cal C}_5} \simeq 1$,  
  the $t/u$-channel processes dominate for the DM pair annihilations  
  when the DM particle is heavier than the Higgs boson.  
In evaluating this process, we may use an effective Lagrangian of the form, 
\bea 
  {\cal L}_{\rm DM-H}^{\rm eff} = \frac{1}{2} \left(\frac{m_W}{v}\right)^2 \frac{{\cal C}_5^2}{m_2} \;
   h \; h \; \overline{\psi_{\rm DM}} \; \psi_{\rm DM} ,
\eea 
which is obtained by integrating $\eta_2$ out, and calculate the DM pair annihilation cross section 
  times relative velocity ($v_{\rm rel}$) as 
\bea 
 \sigma v_{\rm rel} = \frac{1}{64 \pi} \left(\frac{m_W}{v}\right)^4 \left(\frac{{\cal C}_5^2}{m_2}\right)^2 v_{\rm rel}^2
 \equiv \sigma_0 v_{\rm rel}^2. 
\eea
It is well-known that the observed DM relic density is reproduced by $\sigma_0 \sim 1$ pb.  
Since we find $\sigma_0 \sim 0.02$ pb for ${\cal C}_5 \simeq 1$ and $m_2 \simeq M=1$ TeV, 
  we conclude that the observed relic density is not reproduced by the process $\psi_{\rm DM} \psi_{\rm DM} \to h h$.

Next we consider the DM pair annihilation through the $s$-channel Higgs boson exchange 
   when the DM particle is lighter than the Higgs boson.  
Since the coupling between the a pair of DM particles and the Higgs boson is suppressed by $|{\cal C}_1| \ll 1$, 
   an enhancement of the DM annihilation cross section through the Higgs boson resonance 
   is necessary to reproduce the observed relic DM density. 
We evaluate the DM relic abundance by integrating the Boltzmann equation 
\bea
\frac{dY}{dx} = - \frac{xs\langle \sigma v\rangle}{H(m_{\rm DM})} (Y^2-Y^2_{\rm EQ}),
\label{Boltzmanneq}
\eea
where the temperature of the Universe is normalized by the DM mass as $x=m_{\rm DM}/T$, 
 $H(m_{{\rm DM}})$ is the Hubble parameter as $T=m_{\rm DM}$, 
 $Y$ is the yield (the ratio of the DM number density to the entropy density $s$) of the DM particle, 
 $Y_{{\rm EQ}}$ is the yield of the DM in thermal equilibrium, 
 and $\langle \sigma v_{\rm rel} \rangle$ is the thermal average of 
   the DM annihilation cross section times relative velocity for a pair of the DM particles.  
Various quantities in the Boltzmann equation are given as follows: 
\bea
s=\frac{2\pi^2}{45}g_* \frac{m^3_{{\rm DM}}}{x^3}, 
\quad 
H(m_{{\rm DM}}) =\sqrt{\frac{\pi^2}{90} g_*} \frac{m^2_{{\rm DM}}}{M_P}, 
\quad  
sY_{{\rm EQ}} = \frac{g_{{\rm DM}}}{2\pi^2} \frac{m^3_{{\rm DM}}}{x} K_2(x),
\eea 
where $M_P = 2.44 \times 10^{18}$ GeV is the reduced Planck mass, 
 $g_{{\rm DM}}=2$ is the number of degrees of freedom for the DM particle, 
 $g_*$ is the effective total number of degrees of freedom for the particles in thermal equilibrium 
  (in our analysis, we use $g_{*}=86.25$ corresponding to $m_{\rm DM}\simeq m_h/2$ 
  with the Higgs boson mass of 125 GeV), 
  and $K_2$ is the modified Bessel function of the second kind. 
For $m_{\rm DM} \simeq m_h/2 = 62.5$ GeV, a DM pair annihilates into a pair of the SM fermions as  
   $\psi_{{\rm DM}} \psi_{{\rm DM}} \to h \to f \bar{f}$, where $f$ denotes the SM fermions. 
We calculate the cross section for the annihilation process as 
\bea
\sigma(s) =\frac{y^2_{\rm DM}}{16\pi} 
\left[ 3 \left( \frac{m_b}{v} \right)^2 + 3 \left( \frac{m_c}{v} \right)^2 + \left( \frac{m_\tau}{v} \right)^2 \right] 
\frac{\sqrt{s(s - 4m^2_{\rm DM})}}{(s-m_h^2)^2 + m_h^2 \Gamma_h^2}, 
\eea
where $y_{\rm DM}=(m_W/v) |{{\cal C}_1}|$ (see Eq.~(\ref{DMHcoupling})),  
  and we have only considered pairs of bottom, charm and tau for the final states, 
  neglecting the other lighter quarks, and used the following values for the fermion masses 
  at the $Z$-boson mass scale \cite{Bora:2012tx}:  
  $m_b=2.82$ GeV, $m_c=685$ MeV and $m_\tau=1.75$ GeV. 
The total Higgs boson decay width $\Gamma_h$ is given by 
  $\Gamma_h = \Gamma_h^{\rm SM} + \Gamma_h^{\rm new}$, 
  where $\Gamma_h^{\rm SM}=4.07$ MeV \cite{SMGamma_h} 
  is the total Higgs boson decay width in the SM and 
\bea
\Gamma_h^{{\rm new}} =
\left\{
\begin{array}{cc}
0 & m_h < 2 m_{\rm DM} \\
\frac{m_h}{16\pi} \left( 1- \frac{4m_{{\rm DM}}^2}{m_h^2} \right)^{3/2} y_{{\rm DM}}^2 & m_h > 2 m_{\rm DM} \\
\end{array}
\right.,
\eea
  is the partial decay width of the Higgs boson to a DM pair. 
The thermal average of the annihilation cross section is given by
\bea
\langle \sigma v \rangle = (sY_{{\rm EQ}})^{-2} g_{{\rm DM}}^2 
\frac{m_{{\rm DM}}}{64 \pi^4 x} \int_{4m_{{\rm DM}}}^\infty ds 
\hat{\sigma}(s) \sqrt{s} K_1 \left(\frac{x \sqrt{s}}{m_{{\rm DM}}} \right), 
\eea
where $\hat{\sigma}(s) = 2(s-4m_{\rm DM}^2) \sigma(s)$ is the reduced cross section 
 with the total annihilation cross section $\sigma(s)$, 
 and $K_1$ is the modified Bessel function of the first kind. 
We solve the Boltzmann equation numerically 
  and find an asymptotic value of the yield $Y(\infty)$ to obtain the present DM relic density as
\bea
\Omega h^2  = \frac{m_{{\rm DM}}s_0 Y(\infty)}{\rho_c/h^2},
\eea
where $s_0=2890$ cm$^{-3}$ is the entropy density of the present universe,  
 and $\rho_c/h^2=1.05 \times 10^{-5}$ GeV/cm$^3$ is the critical density.

\begin{figure}[t]
\centering
   \includegraphics[width=120mm]{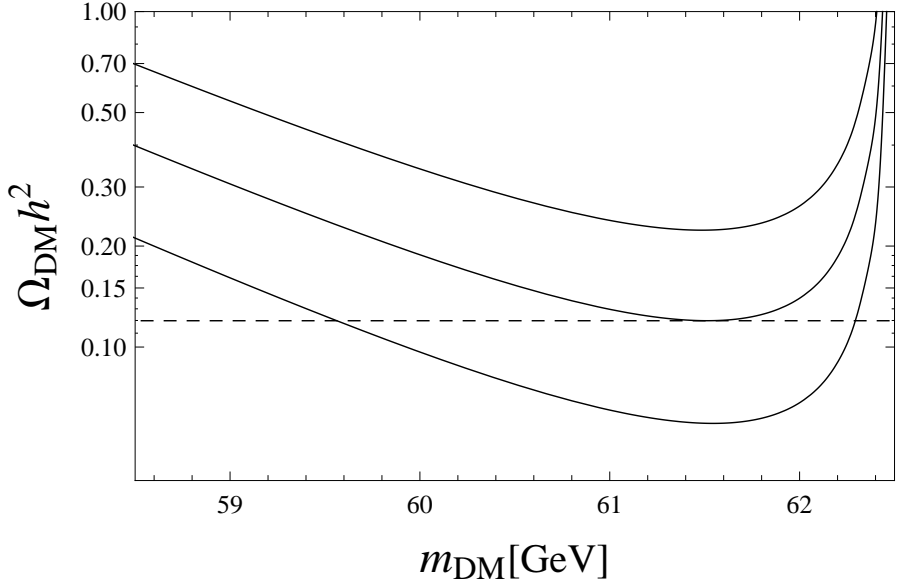}
  \caption{
The DM relic density as a function of the DM mass for various $y_{\rm DM}$ values (solid lines), 
   along with the observed DM density $\Omega_{\rm DM} h^2=0.1198$ (horizontal dashed line)
The three solid lines form top to bottom correspond to $y_{\rm DM}=0.005$, $0.00692$, and $0.01$, respectively.
}
\label{CB}
\end{figure}

In Fig.~\ref{CB} we show the resultant DM relic density as a function of the DM mass 
   for various values of $y_{\rm DM}$.  
The solid lines from top to bottom correspond to  $y_{\rm DM}=0.005$, $0.00692$ and $0.01$, respectively, 
  while the dashed line denotes the observed DM density $\Omega_{{\rm DM}}h^2=0.1198$ 
  from the Planck 2015 result. 
For a fixed $y_{\rm DM}$ value, an intersection of the solid and the dashed lines 
  denotes the DM mass to reproduce the observed DM density. 
We can see that there is a lower bound on $y_{\rm DM} \geq 0.00692$ 
  in order to reproduce the observed DM density. 
        
\begin{figure}[t]
\begin{center}
\includegraphics[width=0.465\textwidth,angle=0,scale=1.05]{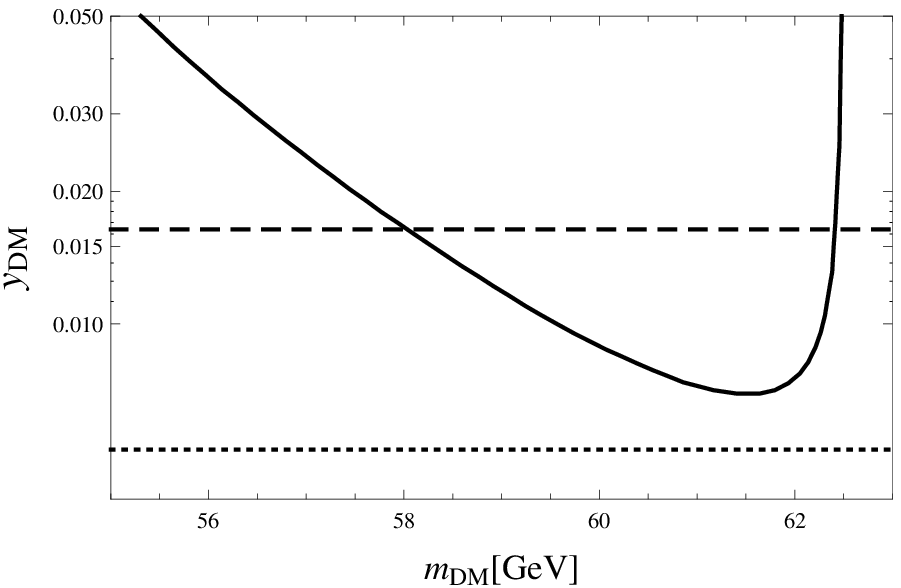}
\hspace{0.1cm}
\includegraphics[width=0.46\textwidth,angle=0,scale=1.05]{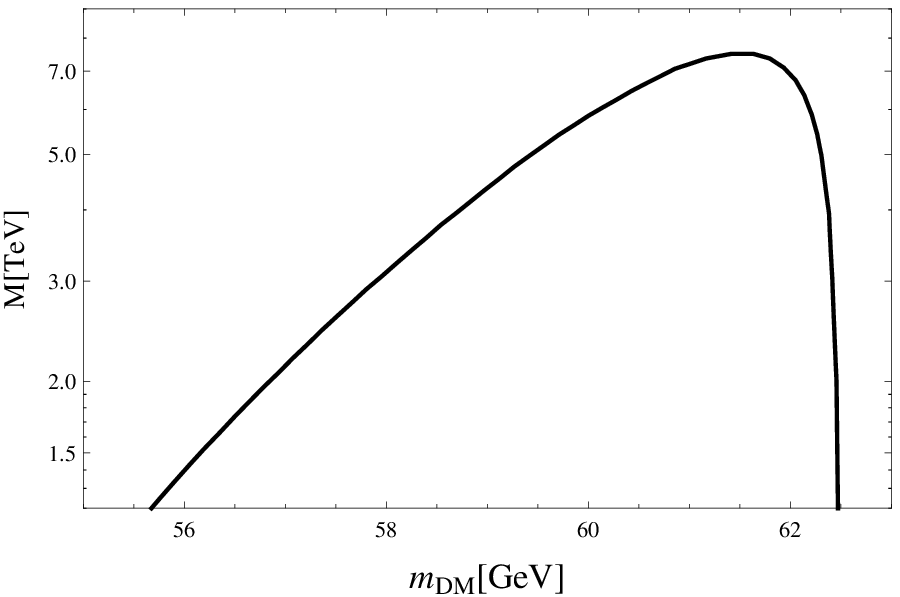}
\end{center}
\caption{
Left panel: $y_{\rm DM}$ as a function of $m_{\rm DM}$ (solid line) 
   along which the observed DM density $\Omega_{\rm DM} h^2=0.1198$ is reproduced.  
Here, the current experimental upper bound from the LUX 2016 result \cite{LUX} and 
   the prospective reach in the future LUX-ZEPLIN DM experiment \cite{LZ}
   are also shown as the dashed and the dotted lines, respectively.  
Right panel: $M$ as a function of $m_{\rm DM}$,  
   along the solid line in the left panel. 
}
\label{Abundance}
\end{figure}

In the left panel of Fig.~\ref{Abundance}, we show $y_{\rm DM}$ as a function of $m_{\rm DM}$ (solid line) 
   along which the observed DM density $\Omega_{\rm DM} h^2=0.1198$ is reproduced.  
Here, the current experimental upper bound from the LUX 2016 result \cite{LUX} 
  and the prospective reach in the future LUX-ZEPLIN DM experiment \cite{LZ} 
  are also shown as the dashed and the dotted lines, respectively, 
  which will be derived in Sec.~\ref{DirectD}.  
In order to satisfy the LUX 2016 constraint, 
   we find the parameter regions such as 
   $58.0 \leq m_{\rm DM}[{\rm GeV}] \leq 62.4$ and $(0.00692 \leq)\; y_{\rm DM} \leq 0.0164$. 
Recall that the Yukawa coupling between the DM particle and the Higgs boson, $y_{\rm DM}=(m_W/v) |{\cal C}_1|$, 
   and the DM mass are determined by the two parameters, $M$ and $m$, 
   from Eqs.~(\ref{MassEigenvalues}), (\ref{u_i}) and (\ref{c_i}). 
Using these formulas, we can express $M$ as a function of $m_{\rm DM}$ along the solid line in the left panel. 
Our result is shown in the right panel. 
Since $M \gg m_W$, the parameter $m$ as a function of $m_{\rm DM}$ is approximately given by 
  $m \simeq M - m_{DM}$. 
Corresponding to the parameter regions of 
  $58.0 \leq m_{\rm DM}[{\rm GeV}] \leq 62.4$ and $(0.00692 \leq)\; y_{\rm DM} \leq 0.0164$,  
  we find $3.14 \leq M[{\rm TeV}] \; (\leq 7.51)$.

\section{Direct Dark Matter Detection} 
\label{DirectD}
A variety of experiments are underway and also planned for directly detecting 
  a dark matter particle through its elastic scattering off with nuclei. 
In this section, we calculate the spin-independent elastic scattering cross section 
  of the DM particle via the Higgs boson exchange to lead to the constraint on 
  the model parameters from the current experimental results.

The spin-independent elastic scattering cross section with nucleon is given by 
\bea
\sigma_{{\rm SI}} = 
\frac{1}{\pi}
\left(
\frac{y_{{\rm DM}}}{v}
\right)^2
\left(
\frac{\mu_{\psi_{{\rm DM}}N}}{m_h^2}
\right)^2
f_N^2
\label{DD}
\eea
where $\mu_{\psi_{{\rm DM}}N} = m_N m_{{\rm DM}}/(m_N+m_{{\rm DM}})$ is 
   the reduced mass of the DM-nucleon system with the nucleon mass $m_N=0.939$ GeV, and 
\bea
f_N = 
\left(
\sum_{q=u,d,s} f_{T_q} + \frac{2}{9}f_{TG}
\right) 
m_N
\eea
   is the nuclear matrix element accounting for the quark and gluon contents of the nucleon. 
In evaluating $f_{T_q}$, we use the results from the lattice QCD simulation \cite{LatQCD}: 
   $f_{T_u} +f_{T_d} \simeq 0.056$ and $|f_{T_s}|\leq 0.08$. 
For conservative analysis, we take $f_{T_s}=0$ in the following.     
Using the trace anomaly formula, $\sum_{q=u,d,s} f_{T_q} + f_{TG}=1$ \cite{TraceAnomaly}, 
   we obtain $f_N^2 \simeq 0.0706 \; m_N^2$ and hence 
\bea
 \sigma_{{\rm SI}} \simeq 
4.47 \times 10^{-7} \; {\rm pb} \times y_{\rm DM}^2 
\label{DD2}
\eea
 for $m_{\rm DM}=m_h/2=62.5$ GeV.

The LUX 2016 result \cite{LUX} currently provides us with the most severe upper bound 
   on the spin-independent cross section, from which we read $\sigma_{\rm SI} \leq 1.2 \times 10^{-10}$ pb 
   for $m_{\rm DM} \simeq 62.5$ GeV. 
 From Eq.~(\ref{DD2}), we find $y_{\rm DM} \leq 0.0164$, 
   which is depicted as the horizontal dashed line in the left  panel of Fig.~\ref{Abundance}. 
The next-generation successor of the LUX experiment, the LUX-ZEPLIN experiment \cite{LZ}, 
   plans to achieve an improvement for the upper bound on the spin-independent cross section 
   by about two orders of magnitude.  
When we apply a conservative search reach to the LUX-ZEPLIN experiment 
   as $\sigma_{\rm SI} \leq 1.2 \times 10^{-11}$ pb  
   (just an order of magnitude improvement from the current  LUX bound),  
   we obtain $y_{\rm DM} \leq 0.00518$. 
This prospective upper bound is shown as the dotted line in the left panel of Fig.~\ref{Abundance}. 
We can see that the present allowed parameter region all covered by the future LUX-ZEPLIN experiment.

\section{Higgs boson mass in effective theory approach}
\label{Hmass}
In this section, we calculate the Higgs boson mass 
  by using a 4-dimensional effective theory approach of the GHU scenario 
  in 5-dimensional Minkowski space, which is developed in Ref.~\cite{GHcondition}.  
In this paper, it has been shown that an effective Higgs quartic coupling 
  derived from the 1-loop effective Higgs potential after integrating out all KK modes 
  coincides with a running Higgs quartic coupling at low energies 
  obtained from the renormalization group (RG) evolution 
  with a vanishing Higgs quartic coupling at the compactification scale (``gauge-Higgs condition" \cite{GHcondition}).  
This vanishing Higgs quartic coupling indicates a restoration of the 5-dimensional gauge invariance   
  at the compactification scale.  
With this approach, we can easily calculate the Higgs quartic coupling at low energies  
  by solving the RG equations, once the particle contents and the mass spectrum 
  of the model below the compactification are defined. 
Assuming that the electroweak symmetry breaking is correctly achieved, 
   the Higgs boson mass is calculated by the Higgs quartic coupling value at the electroweak scale.

There are two scales involved in our RG analysis, namely, the bulk mass $M \simeq m$ 
  and the compactification scale $M_{\rm KK}=1/R$.     
In the following analysis, we ignore the mass splitting among the bulk fermion zero modes  
  and set all of their masses as $M$. 
As we will show in the the following, a hierarchy $M \ll M_{\rm KK}$ is necessary to reproduce 
  the 125 GeV Higgs boson mass, and hence this treatment is justified.

For the renormalization scale smaller than the bulk mass $\mu < M$, 
   all bulk fermions are decoupled and we employ the SM RG equations at two loop level~\cite{RGE}. 
For the three SM gauge couplings $g_i$ ($i=1,2,3$), we have 
\bea
 \frac{d g_i}{d \ln \mu} =
 \frac{b_i}{16 \pi^2} g_i^3 +\frac{g_i^3}{(16\pi^2)^2}
  \left( \sum_{j=1}^3 B_{ij}g_j^2 - C_i y_t^2   \right),
\eea
 where
\bea
b_i = \left(\frac{41}{10},-\frac{19}{6},-7\right),~~~~
 { B_{ij}} =
 \left(
  \begin{array}{ccc}
  \frac{199}{50}& \frac{27}{10}&\frac{44}{5}\\
 \frac{9}{10} & \frac{35}{6}&12 \\
 \frac{11}{10}&\frac{9}{2}&-26
  \end{array}
 \right),  ~~~~
C_i=\left( \frac{17}{10}, \frac{3}{2}, 2 \right). 
\eea 
Here, among the SM Yukawa couplings, we have taken only the top Yukawa coupling ($y_t$) into account. 
The RG equation for the top Yukawa coupling is given by
\bea 
 \frac{d y_t}{d \ln \mu}
 = y_t  \left(
 \frac{1}{16 \pi^2} \beta_t^{(1)} + \frac{1}{(16 \pi^2)^2} \beta_t^{(2)}
 \right), 
\eea
where the one-loop contribution is
\bea
 \beta_t^{(1)} =  \frac{9}{2} y_t^2 -
  \left(
    \frac{17}{20} g_1^2 + \frac{9}{4} g_2^2 + 8 g_3^2
  \right) ,
\eea
while the two-loop contribution is given by 
\bea
\beta_t^{(2)} &=&
 -12 y_t^4 +   \left(
    \frac{393}{80} g_1^2 + \frac{225}{16} g_2^2  + 36 g_3^2
   \right)  y_t^2  \nonumber \\
 &&+ \frac{1187}{600} g_1^4 - \frac{9}{20} g_1^2 g_2^2 +
  \frac{19}{15} g_1^2 g_3^2
  - \frac{23}{4}  g_2^4  + 9  g_2^2 g_3^2  - 108 g_3^4 \nonumber \\
 &&+ \frac{3}{2} \lambda^2 - 6 \lambda y_t^2 .
\eea
The RG equation for the Higgs quartic coupling is given by 
\bea
\frac{d \lambda}{d \ln \mu}
 =   \frac{1}{16 \pi^2} \beta_\lambda^{(1)}
   + \frac{1}{(16 \pi^2)^2}  \beta_\lambda^{(2)},
\eea
with
\bea
 \beta_\lambda^{(1)} &=& 12 \lambda^2 -
 \left(  \frac{9}{5} g_1^2+ 9 g_2^2  \right) \lambda
 + \frac{9}{4}  \left(
 \frac{3}{25} g_1^4 + \frac{2}{5} g_1^2 g_2^2 +g_2^4
 \right) + 12 y_t^2 \lambda  - 12 y_t^4 ,
\eea
and
\bea
  \beta_\lambda^{(2)} &=&
 -78 \lambda^3  + 18 \left( \frac{3}{5} g_1^2 + 3 g_2^2 \right) \lambda^2
 - \left( \frac{73}{8} g_2^4  - \frac{117}{20} g_1^2 g_2^2
 - \frac{1887}{200} g_1^4  \right) \lambda - 3 \lambda y_t^4
 \nonumber \\
 &&+ \frac{305}{8} g_2^6 - \frac{289}{40} g_1^2 g_2^4
 - \frac{1677}{200} g_1^4 g_2^2 - \frac{3411}{1000} g_1^6
 - 64 g_3^2 y_t^4 - \frac{16}{5} g_1^2 y_t^4
 - \frac{9}{2} g_2^4 y_t^2
 \nonumber \\
 && + 10 \lambda \left(
  \frac{17}{20} g_1^2 + \frac{9}{4} g_2^2 + 8 g_3^2 \right) y_t^2
 -\frac{3}{5} g_1^2 \left(\frac{57}{10} g_1^2 - 21 g_2^2 \right)
  y_t^2  - 72 \lambda^2 y_t^2  + 60 y_t^6.
\eea
In solving these RGEs, we use the boundary conditions at the top quark pole mass ($M_t$)
  given in \cite{RGE_Higgs_quartic}: 
\bea 
g_1(M_t)&=&\sqrt{\frac{5}{3}} \left( 0.35761 + 0.00011 (M_t - 173.10) 
   -0.00021  \left( \frac{M_W-80.384}{0.014} \right) \right),  
\nonumber \\
g_2(M_t)&=& 0.64822 + 0.00004 (M_t - 173.10) + 0.00011  \left( \frac{M_W-80.384}{0.014} \right) , 
\nonumber \\
g_3(M_t)&=& 1.1666 + 0.00314 \left(  \frac{\alpha_s-0.1184}{0.0007}   \right) ,
\nonumber \\
y_t(M_t) &=& 0.93558 + 0.0055 (M_t - 173.10) - 0.00042  \left(  \frac{\alpha_s-0.1184}{0.0007}   \right) 
\nonumber \\
  && - 0.00042 \left( \frac{M_W-80.384}{0.014} \right) ,
\nonumber \\
\lambda(M_t) &=&  2 (0.12711 + 0.00206 (m_h - 125.66) - 0.00004 (M_t - 173.10) ) . 
\eea 
We employ $M_W=80.384$ GeV, $\alpha_s =0.1184$, 
   the central value of the combination of Tevatron and LHC measurements of top quark mass 
   $M_t=173.34$ GeV~\cite{top_pole_mass}, and the central value of the updated Higgs boson mass measurement, 
   $m_h=125.09$ GeV from the combined analysis by the ATLAS  and the CMS collaborations~\cite{Higgs_Mass_LHC}.

For the renormalization scale $\mu \geq M$, the SM RG equations are modified 
   in the presence of the bulk fermions.  
In this paper, we take only one-loop corrections from the bulk fermions into account.  
In the presence of a pair of the $SU(3)$ triplet bulk fermions, 
   the beta functions of the $SU(2)$ and $U(1)_Y$ gauge couplings receive new contributions as 
\bea
\Delta b_1=  \Delta b_2 =\frac{2}{3}.    
\eea
The beta functions of the top Yukawa and Higgs quartic couplings are modified as 
\bea 
 \beta_t^{(1)} \to \beta_t^{(1)}  + 2 y_t  |Y_S|^2 ,  \; \; \;
  \beta_{\lambda}^{(1)} \to \beta_{\lambda}^{(1)} + 8 \lambda  |Y_S|^2  - 8 |Y_S|^4,  
\label{LamBeta}
\eea
where $Y_S$ is the universal Yukawa coupling of $\psi$ and $\tilde{\psi}$ with the Higgs doublet 
   in Eq.~(\ref{L_mass}), which obeys the RG equation,  
\bea 
16 \pi^2 \frac{d Y_S}{d \ln \mu} =
  Y_S \left[ 3 y_t^2 - \left( \frac{9}{20}  g_1^2  + \frac{9}{4} g_2^2 \right) 
  + \frac{7}{2} |Y_S|^2 \right]. 
\eea

\begin{figure}[t]
\begin{center}
\includegraphics[width=0.465\textwidth,angle=0,scale=1.05]{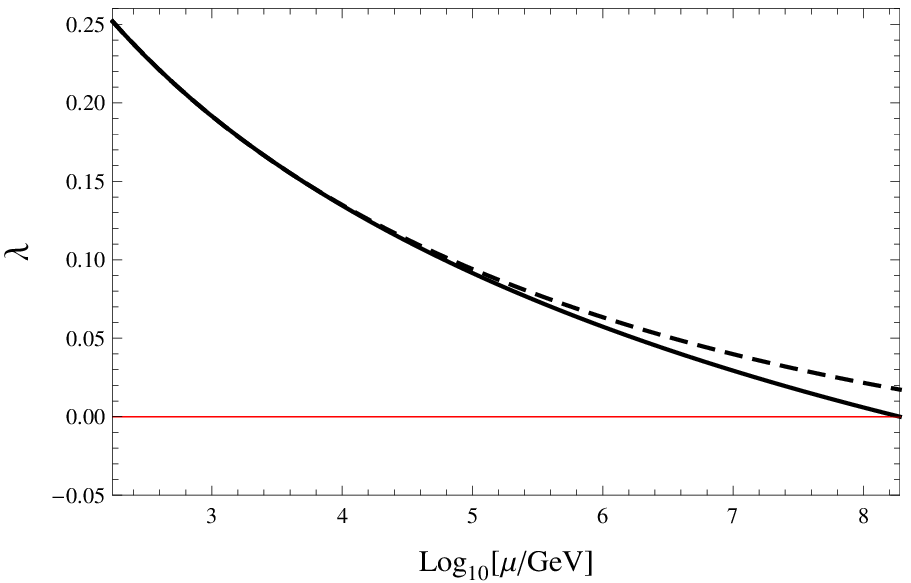}
\hspace{0.1cm}
\includegraphics[width=0.46\textwidth,angle=0,scale=1.05]{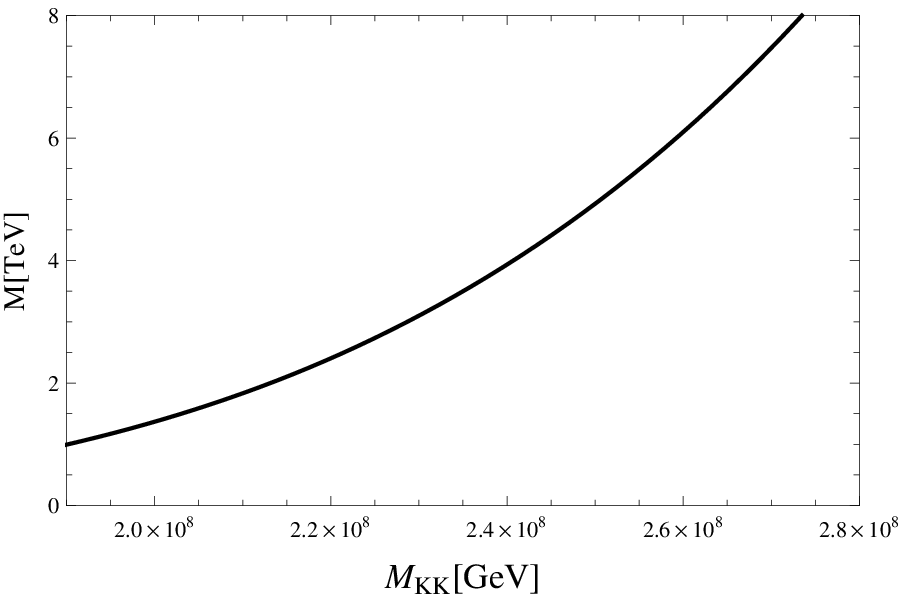}
\end{center}
\caption{
Left panel: 
RG evolution of the Higgs quartic coupling with the bulk mass $M=$1 TeV (solid line), 
   along with the result in the SM (dashed line). 
The compactification scale is found to be  $M_{\rm KK}=1.9 \times 10^8$ GeV,  
   where the gauge-Higgs condition $\lambda(M_{\rm KK})=0$ and 
   $|Y_S(M_{\rm KK})|  = g_2 (M_{\rm KK})/\sqrt{2}$ are satisfied. 
Right panel: the relation between the bulk mass ($M$) and the compactification scale ($M_{\rm KK}$)
                   so as to reproduce the 125 GeV Higgs boson mass. 
}
\label{RGElambda}
\end{figure}

In our RG analysis, we numerically solve the SM RG equations from $M_t$ to $M$, 
    at which the solutions connect with the solutions of the RG equations with the bulk triplet fermions. 
For a fixed $M$ values, we arrange an input $|Y_S(M)|$ value 
   so as to find numerical solutions which satisfy 
   the gauge-Higgs condition and the unification condition 
   between the gauge and Yukawa couplings at the compactification scale, such that 
\bea 
  \lambda(M_{\rm KK})=0, \; \; \;  |Y_S(M_{\rm KK})|  = \frac{g_2 (M_{\rm KK})}{\sqrt{2}} . 
\label{B.C.}  
\eea   
In the left panel of Fig.~\ref{RGElambda} we show the RG evolution of the Higgs quartic coupling (solid line) 
   for $M=1$ TeV, along with the one in the SM (dashed line). 
At $M_{\rm KK}= 1.9 \times 10^8$ GeV, the  boundary condition in Eq.~(\ref{B.C.}) is satisfied. 
For a fixed $M$ value, we numerically find a $M_{\rm KK}$ value. 
Our result for the relation between $M$ and $M_{\rm KK}$ is shown in  the right panel of Fig.~\ref{RGElambda}. 
For $M_{\rm KK} \simeq 10^8$ GeV, the 125 GeV Higgs boson mass is reproduced. 
We obtained the hierarchy $M  \ll M_{\rm KK}$ mentioned above. 
Note that in the absence of the bulk $SU(3)$ triplet fermions, the RG evolution of the Higgs quartic coupling 
  follows the SM one and the compactification scale, at which the quartic coupling becomes zero,  
  is found to be $M_{\rm KK}\simeq 10^{10}$ GeV \cite{RGE}. 
In the presence of the  bulk fermions, the compactification scale is lowered 
  from $M_{\rm KK} \simeq 10^{10}$ GeV to $10^8$ GeV.

\section{Conclusions and discussions}
In this paper, we have proposed a Majorana fermion DM scenario 
   in the context of a 5-dimensional GHU model
   based on the gauge group $SU(3) \times U(1)^\prime$ 
   with a compactification of the 5th dimension on $S^1/Z_2$ orbifold. 
A pair of bulk $SU(3)$ triplet fermions is introduced along with a bulk mass term 
  and a periodic boundary condition.  
The bulk fermions are decomposed into a pair of the $SU(2)$ doublets and 
  a pair of the electric-charge neutral singlets under the SM gauge group of $SU(2) \times U(1)_Y$. 
With Majorana mass terms for the singlets, which are introduced on a brane at an orbifold fixed point in general, 
   the lightest mass eigenstate among the doublet and singlet components serves as a DM candidate. 
 
We have focused on the case that the DM particle is mostly composed of the SM singlet fermions, 
   and have investigated the DM physics. 
In this case, the DM particle communicates with the SM particles through the Higgs boson. 
We have found that an allowed parameter region to reproduce the observed DM density 
   is quite limited and the DM particle mass is to be a vicinity of a half of the Higgs boson mass. 
The allowed region has been found to be further constrained when we take into account 
   the upper limit of the elastic scattering of the DM particle off with the nuclei by the LUX 2016 result. 
We have found that the entire allowed region will be covered by the LUX-ZEPLIN experiment in the near future.

Note that even if the parameter region shown in the left-panel of Fig.~\ref{Abundance} is entirely excluded in the future, 
   our DM scenario can be still viable for the case where the DM particle is mostly a component in the SM $SU(2)$  doublets. 
As mentioned in Sec.~2, the DM particle property in this case is very similar to the Higgsino-like neutralino DM 
   in the MSSM and the observed DM relic abundance is reproduced with the DM mass 
   of around 1 TeV~\cite{ArkaniHamed:2006mb}.  
Since the reduced mass is $\mu_{\psi_{{\rm DM}}N} \simeq m_N$ for $m_{\rm DM} \gg m_N$, 
   we apply Eq.~(\ref{DD2}) for the spin-independent elastic scattering cross section also for the present case.    
However, the limit on $y_{\rm DM}$ is weaker since the experimental upper bound on $\sigma_{\rm SI}$
   for $m_{\rm DM} \sim 1$ TeV is about an order of magnitude 
   higher than the one for $m_{\rm DM} \simeq 62.5$ TeV~\cite{LUX}. 
Furthermore, we can estimate $y_{DM}=(m_W/v)|{\cal C}_1|$ with Eqs.~(\ref{u_i}) and (\ref{c_i}) as
\bea 
    y_{DM} \simeq 2 \left(\frac{m_W}{v}\right) \left(\frac{m_W}{m}\right)
\eea
   for $m \gtrsim M \simeq m_{\rm DM} \sim$ 1 TeV. 
Therefore, in the decoupling limit of the SM singlet components, namely $m \gg M$, 
   the spin-independent elastic scattering cross section is highly suppressed,  
   and therefore the DM particle escapes detection. 
This limit is analogous to the pure Higgsino dark matter in the MSSM.

Employing the effective theoretical approach with the gauge-Higgs condition, 
   we have also studied the RG evolution of Higgs quartic coupling 
   and shown that the observed Higgs mass of 125 GeV is achieved 
   with the compactification scale of around $10^8$ GeV. 
In the presence of the bulk DM multiplets, 
   the compactification scale to reproduce the 125 GeV Higgs boson mass 
   is reduced by about two orders of magnitude from $M_{\rm KK} \simeq 10^{10}$ GeV. 
However, in terms of providing a solution to the gauge hierarchy problem by the GHU scenario, 
   $M_{\rm KK} \simeq 10^8$ GeV is too high for the scenario to be natural. 
In fact, as has been shown in Refs.~\cite{diphoton, MOcolor}, 
   when we introduce a pair of bulk fermions in higher dimensional $SU(3)$ representations 
   such as ${\bf 10}$-plet and ${\bf 15}$-plet, 
   the compactification scale can be as low as ${\cal O}$(1 TeV),  
   while reproducing the 125 GeV Higgs boson mass.  
Hence, toward a natural GHU scenario with a fermion DM, 
   it is worth extending our present model and introducing the bulk DM multiplets  
   in such a higher dimensional representation. 
In this case, we will see that the DM physics investigated in this paper remains 
   almost the same while the Higgs boson mass of 125 GeV can be reproduced 
   with the compactification scale of order 1 TeV~\cite{MOO}.

\section*{Acknowledgments}
S.O. would like to thank the Department of Physics and Astronomy 
  at the University of Alabama for hospitality during her visit. 
She would also like to thank FUSUMA Alumni Association at Yamagata University 
  for travel supports for her visit to the University of Alabama.   
The work of N.O. is supported in part by the United States Department of Energy (Award No.~DE-SC0013680).



\begin{thebibliography}{100}

\bibitem{GH} 
  N.~S.~Manton,
  ``A New Six-Dimensional Approach To The Weinberg-Salam Model,''
  Nucl.\ Phys.\ B {\bf 158}, 141 (1979);
  D.~B.~Fairlie,
  ``Higgs' Fields And The Determination Of The Weinberg Angle,''
  Phys.\ Lett.\ B {\bf 82}, 97 (1979), 
  ``Two Consistent Calculations Of The Weinberg Angle,''
  J.\ Phys.\ G {\bf 5}, L55 (1979);
  Y.~Hosotani,
   ``Dynamical Mass Generation By Compact Extra Dimensions,''
  Phys.\ Lett.\ B {\bf 126}, 309 (1983); 
   ``Dynamical Gauge Symmetry Breaking As The Casimir Effect,''
  Phys.\ Lett.\ B {\bf 129}, 193 (1983); 
  ``DYNAMICS OF NONINTEGRABLE PHASES AND GAUGE SYMMETRY BREAKING,''
  Annals Phys.\  {\bf 190}, 233 (1989).


\bibitem{1loopmass}
  I.~Antoniadis, K.~Benakli and M.~Quiros,
  ``Finite Higgs mass without supersymmetry,''
  New J.\ Phys.\  {\bf 3}, 20 (2001)
  [arXiv:hep-th/0108005]; 
  G.~von Gersdorff, N.~Irges and M.~Quiros,
  ``Bulk and brane radiative effects in gauge theories on orbifolds,''
  Nucl.\ Phys.\ B {\bf 635}, 127 (2002) 
  [arXiv:hep-th/0204223]; 
 R.~Contino, Y.~Nomura and A.~Pomarol,
  ``Higgs as a holographic pseudo-Goldstone boson,''
  Nucl.\ Phys.\ B {\bf 671}, 148 (2003) 
  [arXiv:hep-ph/0306259];  
C.~S.~Lim, N.~Maru and K.~Hasegawa,
  ``Six Dimensional Gauge-Higgs Unification with an Extra Space S**2 and the Hierarchy Problem,''
  J.\ Phys.\ Soc.\ Jap.\  {\bf 77}, 074101 (2008)
  [hep-th/0605180]. 


\bibitem{2loop}
  N.~Maru and T.~Yamashita,
  ``Two-loop calculation of Higgs mass in gauge-Higgs unification: 
  5D  massless QED compactified on $S^1$,''
  Nucl.\ Phys.\ B {\bf 754}, 127 (2006) 
  [arXiv:hep-ph/0603237];  
%
  Y.~Hosotani, N.~Maru, K.~Takenaga and T.~Yamashita,
  ``Two loop finiteness of Higgs mass and potential in the gauge-Higgs unification,''
  Prog.\ Theor.\ Phys.\  {\bf 118}, 1053 (2007) 
  [arXiv:0709.2844 [hep-ph]].
  

  \bibitem{MO}
  N.~Maru and N.~Okada,
  ``Gauge-Higgs Unification at LHC,''
  Phys.\ Rev.\  D {\bf 77}, 055010 (2008) 
   [arXiv:0711.2589 [hep-ph]].


\bibitem{Maru}
  N.~Maru,
  ``Finite Gluon Fusion Amplitude in the Gauge-Higgs Unification,''
  Mod.\ Phys.\ Lett.\  A {\bf 23}, 2737 (2008) 
  [arXiv:0803.0380 [hep-ph]].


\bibitem{diphoton} 
  N.~Maru and N.~Okada,
  ``Diphoton decay excess and 125 GeV Higgs boson in gauge-Higgs unification,''
  Phys.\ Rev.\ D {\bf 87}, no. 9, 095019 (2013)
  [arXiv:1303.5810 [hep-ph]].
  
  
\bibitem{g-2}
  Y.~Adachi, C.~S.~Lim and N.~Maru,
  ``Finite Anomalous Magnetic Moment in the Gauge-Higgs Unification,''
  Phys.\ Rev.\  D {\bf 76}, 075009 (2007); 
  [arXiv:0707.1735 [hep-ph]].
%
  Y.~Adachi, C.~S.~Lim and N.~Maru,
  ``More on the Finiteness of Anomalous Magnetic Moment in the Gauge-Higgs Unification,''
  Phys.\ Rev.\ D {\bf 79}, 075018 (2009)
  [arXiv:0901.2229 [hep-ph]].



\bibitem{EDM}
 Y.~Adachi, C.~S.~Lim and N.~Maru,
  ``Neutron Electric Dipole Moment in the Gauge-Higgs Unification,''
  Phys.\ Rev.\  D {\bf 80}, 055025 (2009) 
  [arXiv:0905.1022 [hep-ph]].


\bibitem{MOcolor}
N.~Maru and N.~Okada,
  ``125 GeV Higgs Boson and TeV Scale Colored Fermions in Gauge-Higgs Unification,''
  arXiv:1310.3348 [hep-ph]; 
%
 J.~Carson and N.~Okada,
  ``125 GeV Higgs boson mass from 5D gauge-Higgs unification,''
  arXiv:1510.03092 [hep-ph].


\bibitem{GHDM}
M.~Regis, M.~Serone and P.~Ullio,
  ``A Dark Matter Candidate from an Extra (Non-Universal) Dimension,''
  JHEP {\bf 0703}, 084 (2007)  
   [hep-ph/0612286]; 
%
G.~Panico, E.~Ponton, J.~Santiago and M.~Serone,
  ``Dark Matter and Electroweak Symmetry Breaking in Models with Warped Extra Dimensions,''
  Phys.\ Rev.\ D {\bf 77}, 115012 (2008) 
   [arXiv:0801.1645 [hep-ph]]; 
%
M.~Carena, A.~D.~Medina, N.~R.~Shah and C.~E.~M.~Wagner,
  ``Gauge-Higgs Unification, Neutrino Masses and Dark Matter in Warped Extra Dimensions,''
  Phys.\ Rev.\ D {\bf 79}, 096010 (2009) 
   [arXiv:0901.0609 [hep-ph]]; 
%
Y.~Hosotani, P.~Ko and M.~Tanaka,
  ``Stable Higgs Bosons as Cold Dark Matter,''
  Phys.\ Lett.\ B {\bf 680}, 179 (2009) 
   [arXiv:0908.0212 [hep-ph]]; 
%
  N.~Haba, S.~Matsumoto, N.~Okada and T.~Yamashita,
  ``Gauge-Higgs Dark Matter,''
  JHEP {\bf 1003}, 064 (2010) 
  [arXiv:0910.3741 [hep-ph]].


\bibitem{SSS} 
  C.~A.~Scrucca, M.~Serone and L.~Silvestrini,
  ``Electroweak symmetry breaking and fermion masses from extra dimensions,''
  Nucl.\ Phys.\ B {\bf 669}, 128 (2003) 
  [hep-ph/0304220].

\bibitem{CCP} 
  G.~Cacciapaglia, C.~Csaki and S.~C.~Park,
  ``Fully radiative electroweak symmetry breaking,''
  JHEP {\bf 0603}, 099 (2006) 
  [hep-ph/0510366].


\bibitem{PDG} 
  C.~Patrignani {\it et al.} [Particle Data Group],
  ``Review of Particle Physics,''
  Chin.\ Phys.\ C {\bf 40}, no. 10, 100001 (2016).



\bibitem{ArkaniHamed:2006mb} 
See, for example,  N.~Arkani-Hamed, A.~Delgado and G.~F.~Giudice,
  ``The Well-tempered neutralino,''
  Nucl.\ Phys.\ B {\bf 741}, 108 (2006)
  [hep-ph/0601041].
  


\bibitem{Planck2015} 
  N.~Aghanim {\it et al.} [Planck Collaboration],
  ``Planck 2015 results. XI. CMB power spectra, likelihoods, and robustness of parameters,''
  Astron.\ Astrophys.\  {\bf 594}, A11 (2016)
  [arXiv:1507.02704 [astro-ph.CO]].


\bibitem{Bora:2012tx} 
  K.~Bora,
  ``Updated values of running quark and lepton masses at GUT scale in SM, 2HDM and MSSM,''
  Horizon {\bf 2} (2013)
  [arXiv:1206.5909 [hep-ph]].



\bibitem{SMGamma_h}
A.~Denner, S.~Heinemeyer, I.~Puljak, D.~Rebuzzi and M.~Spira,
  ``Standard Model Higgs-Boson Branching Ratios with Uncertainties,''
  Eur.\ Phys.\ J.\ C {\bf 71}, 1753 (2011)
  [arXiv:1107.5909 [hep-ph]].


\bibitem{LUX} 
  D.~S.~Akerib {\it et al.} [LUX Collaboration],
  ``Results from a search for dark matter in the complete LUX exposure,''
  Phys.\ Rev.\ Lett.\  {\bf 118}, no. 2, 021303 (2017)
  [arXiv:1608.07648 [astro-ph.CO]].


\bibitem{LZ}
  M.~Szydagis [LUX and LZ Collaborations],
  ``The Present and Future of Searching for Dark Matter with LUX and LZ,''
  arXiv:1611.05525 [astro-ph.CO].


\bibitem{LatQCD} 
  H.~Ohki {\it et al.},
 ``Nucleon sigma term and strange quark content from lattice QCD with exact chiral symmetry,''
  Phys.\ Rev.\ D {\bf 78}, 054502 (2008) [arXiv:0806.4744 [hep-lat]].

\bibitem{TraceAnomaly} 
R.~J.~Crewther,
  ``Nonperturbative evaluation of the anomalies in low-energy theorems,''
  Phys.\ Rev.\ Lett.\  {\bf 28}, 1421 (1972); 
M.~S.~Chanowitz and J.~R.~Ellis,
  ``Canonical Anomalies and Broken Scale Invariance,''
  Phys.\ Lett.\  {\bf 40B}, 397 (1972); 
  ``Canonical Trace Anomalies,''
  Phys.\ Rev.\ D {\bf 7}, 2490 (1973); 
J.~C.~Collins, A.~Duncan and S.~D.~Joglekar,
  ``Trace and Dilatation Anomalies in Gauge Theories,''
  Phys.\ Rev.\ D {\bf 16}, 438 (1977); 
M.~A.~Shifman, A.~I.~Vainshtein and V.~I.~Zakharov,
  ``Remarks on Higgs Boson Interactions with Nucleons,''
  Phys.\ Lett.\  {\bf 78B}, 443 (1978).



\bibitem{GHcondition}
 N.~Haba, S.~Matsumoto, N.~Okada and T.~Yamashita,
  ``Effective theoretical approach of Gauge-Higgs unification model and its phenomenological applications,''
  JHEP {\bf 0602}, 073 (2006) 
  [hep-ph/0511046];
  ``Effective Potential of Higgs Field in Warped Gauge-Higgs Unification,''
  Prog.\ Theor.\ Phys.\  {\bf 120}, 77 (2008) 
  [arXiv:0802.3431 [hep-ph]]. 


\bibitem{RGE}
M.~E.~Machacek and M.~T.~Vaughn,
 ``Two Loop Renormalization Group Equations In A General Quantum Field Theory. 1. Wave Function Renormalization,''
  Nucl.\ Phys.\ B {\bf 222}, 83 (1983);
%
  ``Two Loop Renormalization Group Equations In A General Quantum Field Theory. 2. Yukawa Couplings,''
  Nucl.\ Phys.\ B {\bf 236}, 221 (1984);
%
  ``Two Loop Renormalization Group Equations In A General Quantum Field Theory. 3. Scalar Quartic Couplings,''
   Nucl.\ Phys.\ B {\bf 249}, 70 (1985);
%
C.~Ford, I.~Jack and D.~R.~T.~Jones,
  ``The Standard Model Effective Potential at Two Loops,''
  Nucl.\ Phys.\  {\bf B387} (1992) 373,
  [Erratum-ibid.\  {\bf B504} (1997)  551];
%
H.~Arason, D.~J.~Castano, B.~Keszthelyi, S.~Mikaelian, E.~J.~Piard, P.~Ramond and B.~D.~Wright,
     Phys.\ Rev.\  D {\bf 46}, 3945 (1992);
%
V.~D.~Barger, M.~S.~Berger and P.~Ohmann,
  ``Supersymmetric grand unified theories: Two loop evolution of gauge and
  Yukawa couplings,''
  Phys.\ Rev.\  D {\bf 47}, 1093 (1993);
%
M.~X.~Luo and Y.~Xiao,
  ``Two-loop renormalization group equations in the standard model,''
  Phys.\ Rev.\ Lett.\  {\bf 90}, 011601 (2003).


\bibitem{RGE_Higgs_quartic}
D.~Buttazzo, G.~Degrassi, P.~P.~Giardino, G.~F.~Giudice, F.~Sala, A.~Salvio and A.~Strumia,
  ``Investigating the near-criticality of the Higgs boson,''
  JHEP {\bf 1312}, 089 (2013) 
  [arXiv:1307.3536 [hep-ph]]. 
  

\bibitem{top_pole_mass}
[ATLAS and CDF and CMS and D0 Collaborations],
 ``First combination of Tevatron and LHC measurements of the top-quark mass,''
 arXiv:1403.4427 [hep-ex].


\bibitem{Higgs_Mass_LHC} 
  G.~Aad {\it et al.} [ATLAS and CMS Collaborations],
  ``Combined Measurement of the Higgs Boson Mass in $pp$ Collisions at $\sqrt{s}=7$ and 8 TeV with the ATLAS and CMS Experiments,''
  Phys.\ Rev.\ Lett.\  {\bf 114}, 191803 (2015)
  [arXiv:1503.07589 [hep-ex]].


\bibitem{MOO} 
N.~Maru, N.~Okada and S.~Okada, 
  in preparation.  


\end{thebibliography}
\end{document}